\documentclass[aps,pra,twocolumn,groupedaddress,showpacs,10pt]{revtex4-1}

\usepackage{graphicx}
\usepackage{epstopdf}
\usepackage{amsmath}
\usepackage{verbatim} 
\usepackage{color}   
\usepackage{subfigure}
\usepackage{amssymb}
\usepackage{amsmath}
\usepackage{mathrsfs}
\usepackage{amsfonts}
\usepackage{braket}
\usepackage{appendix}
\DeclareMathAlphabet{\mathpzc}{OT1}{pzc}{m}{it}

\begin{document}

\newcommand{\Cs}{${}^{133}\text{Cs }$}
\newcommand{\Li}{${}^{6}\text{Li }$}
\newcommand{\LiCs}{${}^{6}\text{Li} {}^{133}\text{Cs}$}
\newcommand{\tento}[1]{$\times 10^{#1}$}
\newcommand{\Csma}{$\left|3,+3\right\rangle$}
\newcommand{\Liplus}{$\left|1/2,1/2\right\rangle$}
\newcommand{\Liminus}{$\left|1/2,-1/2\right\rangle$}

\title{Analyzing Feshbach resonances -- A \Li-\Cs case study}

\author{R. Pires}
\author{M. Repp}
\author{J. Ulmanis}
\author{E. D. Kuhnle}
\author{M. Weidem\"uller}
\email[]{weidemueller@uni-heidelberg.de}
\affiliation{Physikalisches Institut, Ruprecht-Karls Universit\"{a}t Heidelberg, Im Neuenheimer Feld 226, 69120 Heidelberg, Germany}

\author{T. G. Tiecke}
\email[]{tiecke@physics.harvard.edu}
\affiliation{Department of Physics, Harvard University, Cambridge, Massachusetts, 02138, USA}

\author{Chris H. Greene}
\email[]{chgreene@purdue.edu}
\affiliation{Department of Physics, Purdue University, West Lafayette, Indiana, 47907-2036, USA }

\author{Brandon P. Ruzic}
\author{John L. Bohn}
\affiliation{JILA, University of Colorado and National Institute of Standards and Technology, Boulder, Colorado 80309-0440, USA }

\author{E. Tiemann}
\email[]{tiemann@iqo.uni-hannover.de}
\affiliation{Institut f\"{u}r Quantenoptik, Leibniz Universit\"{a}t Hannover, Welfengarten 1,  30167 Hannover, Germany}

\date{\today}

\begin{abstract}
We provide a comprehensive comparison of a coupled channels calculation, the asymptotic bound state model (ABM), and the multichannel quantum defect theory (MQDT). Quantitative results for \Li-\Cs are presented and compared to previously measured \Li-\Cs Feshbach resonances (FRs) [M. Repp et al., Phys. Rev. A 87 010701(R) (2013)]. We demonstrate how the accuracy of the ABM can be stepwise improved  by including magnetic dipole-dipole 
interactions and coupling to a non-dominant virtual state. We present a MQDT calculation, where magnetic dipole-dipole and second order spin-orbit interactions are included. A frame transformation formalism is introduced, which allows the assignment of measured FRs with only three parameters. All three models achieve a total rms error of $<1\,\mathrm{G}$ on the observed FRs. We critically compare the different models 
in view of the accuracy for the description of FRs and the required input parameters for the calculations.

\end{abstract}


\maketitle

\section{Introduction}
One of the outstanding properties in the field of atomic physics is the ability to control interatomic interactions  using magnetically tunable Feshbach resonances (FRs) \cite{Chin2010}. They allow to address key problems in several fields of physics. For example, in order to explore molecular physics, one can create deeply bound molecules via Feshbach association \cite{Regal2003,Herbig2003}, followed by stimulated Raman adiabatic passage \cite{Ni2008,Danzl2008,Lang2008}. Such molecules can be used for the study of molecular structure, ultracold chemistry, and precision tests of fundamental laws of nature \cite{Carr2009}. Another example for the use of FRs is the study of the BEC-BCS crossover regime \cite{Regal2004,Bartenstein2004,Zwierlein2004} and the transition from weak to strong interactions \cite{OHara2002,Donley2001} in atomic many body physics. The tunability of the two-body scattering length is applied for the creation of Efimov trimers \cite{Braaten2007} in order to investigate few-body physics.

For the study of the above mentioned phenomena, precise knowledge of the field-dependent scattering lengths is essential. This can be obtained via a straightforward numerical coupled channels calculation (CC), which often employs a large number of channels $N$. As the time for the matrix operation required to solve this problem is on the order of $N^{3}$ \cite{Croft2011}, such a calculation can be computationally expensive. However, sufficient insight can be gained by applying models that approximately describe the scattering properties, while reducing the computational effort enormously. Two such models have been proven as powerful alternatives. 

One of these models is the asymptotic bound state model (ABM) \cite{Tiecke2010a,Wille2008}, which uses only the bound states close to the asymptote to describe observables like FRs and the scattering length, removing the computation of the spatial part of the Schr\"odinger equation and the continuum of scattering states. A second approach to calculate scattering observables is the multichannel quantum defect theory (MQDT) \cite{Greene1979,Greene1982}, which uses the separation of length and energy scales to facilitate the calculation. 

Even though Feshbach resonances have been extensively reviewed in Ref.\cite{Chin2010}, the literature is currently lacking a detailed juxtaposition of the aforementioned models. The goal of the present paper is to fill this gap by comprehensively comparing the approaches of CC calculation, ABM, and MQDT and by providing quantitative results based on the example of the \Li-\Cs system.

The reason for choosing this specific atom combination is the special role it will exhibit for the investigation of the above mentioned applications of FRs. For example, with the largest permanent electric dipole moment among all alkali-atom combinations of 5.5 Debye \cite{Aymar2005,Deiglmayr2010}, LiCs molecules in their rovibrational ground state \cite{Deiglmayr2008} are a unique candidate for the study of dipolar quantum gases \cite{Pupillo2008a}. Additionally, the large mass ratio of $m_{Cs}/m_{Li}\approx 22$ results in a very favorable Efimov scaling factor of 4.88 \cite{DIncao2006}, thus enabling the observation of a series of Efimov resonances~\cite{Pires2014,Tung2014}. Moreover, the system is also an excellent candidate for the study of polaron physics \cite{Tempere2009,Cucchietti2006}, because one resonance overlaps with a zero crossing of the \Cs scattering length, which allows for a strong coupling of a \Li impurity to a noninteracting Cs BEC.

We have recently reported on the observation of 19 intraspecies \Li-\Cs $s$- and $p$-wave FRs, which have been accurately assigned via a CC calculation \cite{Repp2013} with a root mean square (rms) deviation $\delta B^{rms}$ (for a definition see Eq.~\ref{eq:rms}) of 39 mG for the field positions of the observed resonances. An application of the crudest version of the ABM with six free fit parameters, similar to the one done in Ref.~\cite{Repp2013}, yields $\delta B^{rms}=877$~mG. However, leaving all six parameters as free parameters in the fit yields unphysical fit values because the parameters are significantly correlated. Therefore, we demonstrate how this fit can be improved 
 by minimizing the amount of free fit parameters and by including 
 magnetic dipole-dipole interaction, yielding a slightly increased $\delta B^{rms}=965$ mG but parameters that are physically consistent and are coming close to those derived in the CC analysis. 

The \Li-\Cs combination is a good system for the illustration of extensions to the ABM, 
because its small reduced mass leads to a large spacing between vibrational states. Therefore, only the least bound states need to be included, which keeps the number of parameters low, and minimizes the computational effort. Other systems with higher reduced mass would require a larger number $n$ of bound states, which results in $2n+n^{2}$ fit parameters ($2n$ bound states in singlet and triplet potentials and $n^2$ respective overlap parameters). For example in Rb-Cs at least five vibrational levels have to be included. The required 35 parameter fit to the observed resonances is asking for an appropriate number of observations if no further theoretical input is available.

We additionally apply the dressed ABM, which includes the coupling of the bound molecular state to the scattering state of the incoming atoms \cite{Tiecke2010a}, to improve the agreement with experimental FR positions in the \Li-\Cs~system even further. The application of this model is not straightforward due to a subtlety in the \LiCs~triplet potential. A virtual state, which is close to the atomic threshold, is not resonant enough to dominate the scattering behavior in the open channel. Therefore, neither the limiting case where a bound state dominates \cite{Tiecke2010a}, nor the case where only the virtual state dictates the behavior \cite{Marcelis2004} is applicable. We will bridge this gap by demonstrating a phenomenological method that includes both effects, leading to a convincing description of the observed FRs with a rms deviation of 263 mG.

Unlike the ABM, the MQDT handles the spatial part of the scattering problem at large separation $R$ explicitly, and the formalism does not differentiate between dominating bound or virtual states. Thus, the latest version of the MQDT as described in Ref.~\cite{Ruzic2013} can be directly applied without extension, resulting in a rms deviation of 40 mG. Besides giving the results for the \Li-\Cs~case, we demonstrate how a frame transformation (FT) in a MQDT ansatz can be applied to a system where no accurate potentials and only experimental data for FR positions are available, in order to assign these resonances and predict other resonance positions. The rms deviation of the FT approximation for the \Li-\Cs~system becomes 48 mG.

This paper is organized as follows. In Sect.~\ref{sec:nutshell} we explain the basic approach and the underlying assumptions of the three models to the scattering problem. The results of CC calculation are given in Sect.~\ref{sec:channels}. Sect.~\ref{sec:ABM} demonstrates how the ABM can be stepwise extended to predict the position of the \Li-\Cs FRs more accurately. In Sect.~\ref{sec:MQDT} we discuss the results of the MQDT calculation and finally, in Sect.~\ref{sec:Conclusion} we provide the quantitative comparison of the models and summarize our results.

\section{Approaches to the scattering problem in a nutshell}
\label{sec:nutshell}

The scattering process of two colliding atoms can be described by the following Hamiltonian \cite{Stoof1988}:
\begin{equation}
\label{eq:Hamiltonian}
H=T+V+H_{\text{hf}}+H_{\text{Z}}+H_{\text{dd}},
\end{equation}
where $T=-\hbar^{2}\Delta^{2}/(2\mu)$ denotes the relative kinetic energy term, with reduced mass $\mu$, and $V$ denotes the potential energy curves. The hyperfine energy operator
\begin{equation}
\label{eq:Hyperfineterm}
H_{\text{hf}}=\sum_{\beta=A,B} \alpha_{\beta}(R) \vec{s}_{\beta}\cdot \vec{i}_{\beta}/\hbar^{2},
\end{equation}
contains the electronic and nuclear spin operators $ \vec{s}$ and $\vec{i}$, respectively, and the summation is performed over the two atoms A and B. In the limit of large separations, the functions $\alpha_{\beta}(R)$, which depend on the internuclear separation $R$, approach the atomic hyperfine constant $\mathrm{a}_\mathrm{hf}$. The Zeeman interaction is given by
\begin{equation}
\label{eq:ZeemanTerm}
H_{\text{Z}}=\sum_{\beta=A,B} (g_{s,\beta}s_{z,\beta}+g_{i,\beta}i_{z,\beta})\mu_{B}B/ \hbar,
\end{equation}
where $g_{s}$ ($g_{i}$) is the electron (nuclear) g-factor, with respect to the Bohr magneton $\mu_{B}$ (see Ref.~\cite{Arimondo1977}). $H_{\text{dd}}$ is the Hamiltonian describing direct magnetic spin-spin, as well as second-order spin-orbit interactions, which causes for example the observed splitting of $p$-wave resonances \cite{Repp2013}. It can be given in its effective form \cite{Strauss}:
\begin{equation}
\label{eq:Vdipole}
V_\mathrm{dip}(R)=\frac{2}{3} \lambda(R)(3S_Z^2-S^2),
\end{equation} 
where $S_Z$ is the total electron spin $S$ projected onto the molecular axis. The function
\begin{equation}
\label{eq:lamda}
\lambda(R) = -\frac{3}{4}
\alpha^2\left(\frac{1}{R^3}+a_\mathrm{SO}
\exp{\left(-bR\right)}\right),
\end{equation}
is given in atomic units with $\alpha$ the universal fine structure constant. Because the parameters $b$ and $a_\mathrm{SO}$ for the assumed effective functional form of the second order spin-orbit interaction are not available in the literature, they become fitting parameters in the following discussion. For binary collisions of alkali atoms, the total spin $S=s_{A}+s_{B}$ can only be 0 or 1. Therefore, the interatomic interaction $V=P_{0}V_{0}+P_{1}V_{1}$ is projected onto the singlet ($V_{S=0}$) and triplet ($V_{S=1}$) components by the projection operators $P_{0}$ and $P_{1}$, respectively, and additionally contains a centrifugal term from the separation of $T$ in radial and angular motion. The manifold of different internal states connected to the Hamiltonian of Eq.~\ref{eq:Hamiltonian} defines a number of channels for a given space fixed projection M of the total angular momentum of the system. Unless otherwise stated, the coordinates connected to spin and angular momentum are characterized by use of an appropriate basis set like in Hund's coupling case (e) for an atom pair AB: 
\begin{equation}
\label{channelsdefinition}
|\chi \rangle \equiv|(s_A,i_A)f_A,m_A;(s_B,i_B)f_B,m_B,l,M>,
\end{equation}
 where the electron spin $s$ couples with the nuclear spin $i$ to the atomic angular momentum $f$ with its projection $m$ on the space fixed axis. $l$ is the quantum number of the overall rotation of the atom pair. The basis vectors in Eq.~(\ref{channelsdefinition}) can be interpreted in two ways, namely for the field-free case, where $f_A$ and $f_B$ are good quantum numbers or in a magnetic field where the pair is build up by the eigenvectors of the Breit-Rabi formula and $f_A$ and $f_B$ are approximate quantum numbers to label the corresponding eigenvector. The channel with the same spin state as the incoming atoms, for which we want to find the FRs, is called entrance channel. Those channels with an asymptotic ($R \rightarrow \infty$) energy  larger than that of the entrance channel are called closed channels, all others are referred to as open channels.

In principle, it is impossible to solve the corresponding Schr\"odinger equation without any approximations due to the fact that an infinite number of coupled channel equations, from an infinite number of basis states, are involved. In the following, we will give a general description of three different models to overcome this difficulty in order to obtain an accurate description of resonance positions, using the \Li-\Cs \ system as an example.

\subsection{Coupled Channels Calculation}
\label{sec:NutshellCC}

The coupled channels calculation is a numerical approach to solve the Schr\"odinger equation resulting from the Hamiltonian of Eq.~\ref{eq:Hamiltonian}. For bound states, $R$ is represented on a grid and the resulting matrix is diagonalized, while for scattering solutions, the logarithmic derivative of the wave function is propagated in discrete steps with optimized step size from low $R$ to large $R$, from which the phase shift is determined by comparing with asymptotic wave functions. To calculate bound states, the wave functions at small separations $R_{in}$ and large separations $R_{out}$ (up to 10 000 $a_{\mathrm{0}}$ for the weakest bound levels, where $a_{\mathrm{0}}$ represents the Bohr radius) are set to zero as boundary conditions. This is equivalent to adding an infinitely high potential wall at $R_{in}$ and  $R_{out}$, resulting in discretized continuum states, often referred to as box states. As this leads to shifts of the calculated resonance states, the size of the modeled box potential will be increased for achieving the desired accuracy. 

Furthermore, in order to obtain a finite number of equations, the basis set is truncated, which is usually called close-coupling calculation. The attribute "close" refers to the fact that only states which are "close" in energy to each other, are retained. In the present approach the truncation is only in the space spanned by the rotational quantum number $l$ and naturally by using only the two molecular ground states X$^1\Sigma ^+$ and a$^3\Sigma ^+$. We span all spin channels allowed by given $s_A$ and $s_B$ as well as $i_A$ and $i_B$ and the chosen space fixed projection $M$ of the total molecular angular momentum. The coupling to higher electronic states exists but is weak and to some degree contained in $H_{dd}$. For collisions of alkali atoms in the ground state at ultracold temperatures, only a limited number of partial waves $l$ has to be included, owing to the small collision energy.

Performing the numerical procedure for a fine grid of magnetic fields yields the field dependent collisional properties, e.g. scattering lengths, collisional cross sections and collision rates. The procedure as we apply it, is specified in Refs.~ \cite{Marzok2009,Schuster2012}, and our results for \Li-\Cs are provided in Sect.~\ref{sec:channels}.

\subsection{Asymptotic Bound State Model}
\label{sec:NutshellABM}

The ABM simplifies the calculation of the coupled Schr\"odinger equations by replacing the kinetic energy term and the interatomic potentials in Eq~(\ref{eq:Hamiltonian}) by their bound-state energies as adjustable parameters for describing the observed FRs, and neglecting the scattering continuum \cite{Tiecke2010a,Wille2008}. Therefore, neither accurate potentials, which are often not available, nor numerical integration of the spatial Schr\"odinger equation are needed. Solving the eigenvalue problem with the approximate Hamiltonian reduces to a simple matrix diagonalization of low dimension, which is the major benefit of the model. The ABM \cite{Tiecke2010a} has been introduced in Ref.~\cite{Wille2008} and builds upon a model by Moerdijk et al. \cite{Moerdijk1995}. Since then it has been extended to include various physical phenomena which has been applied to describe Feshbach resonances in many systems \cite{Wille2008,Li2008,Voigt2009,Deh2010,Knoop2011,Goosen2010,Tscherbul2010,Repp2013,Park2012,Goosen2011}. The ABM model is explained in detail in Ref.  \cite{Tiecke2010a} and here we present a summary and describe various extensions to the model. 

We begin by considering zero-energy collisions ($E_{kin}=0$) and restrict ourselves to $s$-wave collisions where $\left\langle H_\mathrm{dd}\right\rangle =0$. The model introduced by Moerdijk et al. \cite{Moerdijk1995} neglected coupling of the singlet and triplet states reducing the Hamiltonian (\ref{eq:Hamiltonian}) to: $H=\epsilon_{0,1}+H_\mathrm{hf}^+ + H_\mathrm{Z}$ where $\epsilon_{0,1}$ represent the singlet and triplet bound state energies and $H_\mathrm{hf}^+$ is the part of the hyperfine interaction which does not couple singlet and triplet states. This is a valid approximation for the special case that the spacing between the singlet and triplet energies is larger than the hyperfine energy. 
In the ABM, the full hyperfine interaction $H=\epsilon_{0,1}+H_\mathrm{hf} + H_\mathrm{Z}$ is included, which generalizes the Moerdijk model to systems with arbitrary bound state energies, and the singlet-triplet coupling is characterized by the overlap integral $\zeta_{l}=\left\langle \Psi^{l}_{S=0}|\Psi^{l}_{S=1}\right\rangle $ of the singlet ($|\Psi^{l}_{S=0}\rangle $) and triplet  ($|\Psi^{l}_{S=1}\rangle $)  wavefunctions times the nondiagonal part of the Hamiltonian.

In the ABM the Hilbert space consists of only bound states and no scattering states. Therefore, the calculation includes only the basis states 
\begin{equation}
|\sigma \rangle\equiv |S M_{S} m_{iA} m_{iB} v_{n,S} l> 
\label{eq:NewABM_basis}
\end{equation} of pure electon spin states $S=0$ or $S=1$, which will be related to the respective channels (see Eq.~\eqref{channelsdefinition}) at a later stage for a pair of vibrational levels of the singlet and triplet state together. $M_{S},m_{iA}$ and $m_{iB}$ are the projections onto the space fixed axis of the operators $S,i_{A}$ and $i_{B}$, respectively, and $v_{n,S}$ is the $n$-th vibrational state in the $S=1$ or $S=0$ state. 
The FRs are found at the magnetic fields for which an eigenstate exists at the energy of the incoming atom pair at that field. This condition corresponds to $E_{kin}=0$. Additionally, if $\left\langle H_\mathrm{dd}\right\rangle$ is small enough to be neglected, the Hamiltonian (\ref{eq:Hamiltonian}) is diagonal in the partial wave quantum number $l$. As a result, the only parameters needed for the calculation of the FRs in each partial wave $l$ are the energies of the bound states $\epsilon^{l}_{S}$ of the singlet ($S=0$) and triplet ($S=1$) potentials and their wavefunction overlap  $\zeta_{l}$. In fact, only a small number of such states has to be taken into consideration, because the FRs usually arise from the least bound states close to the asymptote. The energies $\epsilon^{l}_{S}$ and the overlap parameters $\zeta_{l}$ are the free parameters of the ABM and are typically obtained by fitting to experimentally observed FRs. 

The resulting Schr\"odinger equation can be written in the form of a $N \times N$ matrix, denoted by $\underline{M}_{ABM}$, where $N$ is determined by the number of spin channels and the number of selected vibrational states; N is on the order of a few tens. The diagonalization of this matrix for different fields provides the molecular energies as a function of the magnetic field. A comparison of this function to the energy sum of the two atoms yields the magnetic fields, at which the energies of bound-states and incoming free atoms are degenerate, thus marking the position of the FRs, as depicted in Fig.~\ref{fig:EnergyLevels}. 

Close to a $s$-wave resonance, the molecular state --and therefore the resonance position-- is shifted due to coupling to the scattering states of the open channel. These states are continuum states and hence not included in the ABM model as described above. However, in some systems, the coupling has such a severe effect on the resonance position that it cannot be neglected, but it can be approximated by the coupling of the resonant molecular state to the least bound state of the open channel\cite{Tiecke2010a}, which requires assigning the bound-states of $\underline{M}_{ABM}$ to the scattering channels.  

For this purpose, a rotation of the basis of  $\underline{M}_{ABM}$ is performed: from the $|\sigma\rangle$ basis (constructed for a singlet and triplet vibrational level) to the basis formed by the eigenvectors of $H_\mathrm{hf}+H_\mathrm{Z}$ at the desired magnetic field (see Eq.~\eqref{channelsdefinition}). This can be ordered in the block matrix 
\begin{equation}
\label{eq:ABMmatrix}
\underline{M'}_{ABM}=\left(\begin{array}{cc}
\mathcal{H}_{PP} & \mathcal{H}_{PQ} \\ 
\mathcal{H}_{QP} & \mathcal{H}_{QQ}
\end{array}  \right),
 \end{equation}
where the index $P$ ($Q$) stands for the spin states which are associated with an open (closed) channel and might include possible $l$ partial waves. A diagonalization of the submatrix $\mathcal{H}_{QQ}$ provides the bare molecular energies $\epsilon_{Q}$, which are the energies of the molecular state when no coupling to the open channel bound state occurs. Typically, only one of these states is the resonant state which causes the FR under consideration.

With the assumption that near a resonance the system can be described in a two channel picture, with one incoming, open channel and one resonant, closed channel, the total $S$-matrix of the scattering problem in the open channel can be written in the simple form of Eq.~(22) in Ref.~\cite{Tiecke2010a} at energy E with wave vector amplitude $|k|=(2\mu |E|)^{1/2}/\hbar$. 

For the calculation of the Feshbach resonances, which are given by the poles of the scattering matrix, the complex energy shift $\mathcal{A}(E)$ locating the pole needs to be estimated. Depending on whether a bound state or a virtual state dominates the scattering behavior, different expressions have to be used for $\mathcal{A}(E)$.
E.g. for $^{40}$K-$^{40}$K collisions a real bound state of the open channel (with wavenumber $k_{p}=i\kappa_{bs}$ with $\kappa_{bs}>0$) occurs close to resonance resulting in a large positive background scattering length. 
In this case $\mathcal{A}(E)$ is given by \cite{Tiecke2010a}
\begin{equation}
\label{eq:BScomplexShift}
 \mathcal{A}(E)=\frac{\mu}{\hbar^{2}} \frac{-iA}{\kappa_{bs}(k-i\kappa_{bs})},
\end{equation}
 where $\kappa_{bs}$ is the wavevector associated with the bare energy of the open channel $\epsilon_{bs}<0$, which is found on the diagonal of the submatrix $\mathcal{H}_{PP}$ in Eq.~\eqref{eq:ABMmatrix}. The coupling term $A$ is the square of the appropriate off-diagonal matrix element in $\mathcal{H}_{PQ}$ between the $P$-channel and the resonant $Q$-channel, after the $Q$ subspace has been diagonalized and $\underline{M'}_{ABM}$ has been transformed to the eigenvector of $Q$ space. This procedure allows for a prediction of the resonance width (imaginary part of $\mathcal{A}(E)$) and shift (real part of $\mathcal{A}(E)$) arising from coupling to the continuum without additional parameters. Using the $S$-matrix, the scattering properties around the resonance can be derived. In the present case we consider only the positions of Feshbach resonances; these will appear at $E=0$ and $k=0$ for a magnetic field where the bare molecular energy satisfies $\epsilon_Q=-(\mu/\hbar^2) A/ \kappa_{bs}^2=-A/2|\epsilon_{bs}|$. 
 
A virtual state, which is also often referred to as an anti-bound state ($k_{p}=-i\kappa_{vs}$ and $\kappa_{vs}>0$  \cite{Marcelis2004}) results in a large negative background scattering length. The $^{6}$Li$-^{6}$Li \cite{Abraham1997} and $^{133}$Cs$-^{133}$Cs \cite{Leo2000} systems are excellent examples for a system with a dominating virtual state.
In this scenario, the complex energy shift is given by \cite{Marcelis2004}
\begin{equation}
\label{eq:VScomplexShift}
 \mathcal{A}(E)=\frac{\mu}{\hbar^{2}} \frac{-iA_{vs}}{\kappa_{vs}(k+i\kappa_{vs})},
\end{equation}
where the coupling between virtual and bound state $A_{vs}$ enters as new parameter, while $\kappa_{vs}$ can be estimated from the van der Waals range $r_{0}$ via $a_{bg}=r_{0}-1/\kappa_{vs}$. To find the position of Feshbach resonances one has to look for magnetic fields where the binding energy of the bare molecular state $\epsilon_Q=+(\mu/\hbar^2) A_{vs}/\kappa_{vs}^2$. 

To calculate the background scattering length of the desired open channel $a_{bg}$, one requires the singlet ($a_{S}$) and triplet ($a_{T}$) background scattering lengths, as well as a decomposition of the ABM matrix eigenstates into singlet and triplet components. $a_{S}$ and $a_{T}$ can be estimated via the accumulated phase method, which employs a numerical calculation of the singlet and triplet wave functions from the asymptotic form of the inter-atomic potential $V_{as}$, using only the van der Waals tail plus adding the centrifugal barrier and the bound state energies. This procedure is described in Refs. \cite{Tiecke2010a,Verhaar2009}. 
Obtaining the poles of the $S$-matrix for a system in which the virtual state dominates the scattering behavior has been utilized in Ref. \cite{Park2012} to explain FRs in a NaK mixture using the ABM.

The \Li-\Cs system, however, is in an intermediate regime, where both the bound state and the virtual state in the open channel are required to describe the FR positions. In Sect.~\ref{sec:ABM} we demonstrate an extension of the existing models, that starts from the virtual state description, but includes the coupling to the bound state in a phenomenological way.

\subsection{Multichannel Quantum Defect Theory}

The MQDT uses a separation of the solution to the Schr\"odinger equation into a long-range and a short-range part. It is based on a model by Seaton \cite{Seaton1983}, which was originally introduced to describe the properties of an electron in the field of an ion. However, it has been generalized in Refs.~\cite{Greene1979,Greene1982} and can now be applied to a variety of collisional partners, with all sorts of interaction potentials (see Ref.~\cite{Croft2011} and references therein). For example, it has been applied successfully to various neutral atom pairs \cite{Burke1998,Gao2005,Raoult2004,Mies2000,Julienne1989,Gao1998,Gao2001}, and can, in general, be used for all alkali atom combinations without adaptation. The most recent modification improves the model for an accurate description of higher partial waves \cite{Ruzic2013}.

The main benefit of the model stems from the separate treatment of the long-range part of the scattering problem, where the van der Waals interaction dominates over exchange interactions and higher order terms. It can be solved accurately using the Milne phase amplitude method (see Ref. \cite{Burke1998} and references therein). This results in a linearly independent pair of functions $(f^{0},g^{0})$, referred to as base pair, which are smooth and analytic functions of energy. In the short-range part, the coupled Schr\"odinger equation at energy $E$ is numerically integrated outwards to a radius $R_{lr}$ on the order of a few tens of atomic units (typically 30 $a_{0}$), beyond which the exchange interaction is negligible. At $R_{lr}$ it is then connected to the long-range part of the solution.

The calculation incorporates only those channels which have a non-negligible effect on the scattering behavior of the system by truncating the basis set of Eq.~\eqref{channelsdefinition} in the same manner as for the CC model. The solution is given by the square matrix $\underline{M}(R)$, which contains the independent solutions of each channel in its columns. Beyond $R_{lr}$, $\underline{M}(R)$ can be given as superposition of the base pair:
\begin{equation}
\label{eq:Mmatrix}
\underline{M}(R)=\underline{f}^{0}(R)-\underline{g}^{0}  \underline{K}^{sr},
\end{equation}
where $\underline{f}^{0} \ \mathrm{and} \ \underline{g}^{0}$  are diagonal matrices which contain the base pair evaluated at the appropriate channel energies $\epsilon_{i}=E-E_{i}$. In this notation $E_{i}$ is the energy of the asymptote of channel $i$. The short-range reaction matrix $\underline{K}^{sr}$ contains all the system specific information for the scattering behavior at low energies. Besides the short-range reaction matrix, one needs four coefficients in order to construct the $S$-matrix, which delivers the physical observables. Detailed instructions on how to obtain these coefficients, which are often noted as $A$, $\mathcal{G}$, $\gamma$ and $\eta$, are given in Refs.~\cite{Ruzic2013,Burke1998,Burke1999}.

The next level of simplification of the MQDT is the assumption that  $\underline{K}^{sr}$ depends only very weakly on energy. Thus it only needs to be calculated for a few energies, and can then be interpolated between these values. In the best case, a $\underline{K}^{sr}$ matrix which is only calculated for one energy (typically close to threshold) and at zero magnetic field can be utilized to describe the scattering properties over a wide range of energies and magnetic fields. However, for obtaining  $\underline{K}^{sr}$, one still has to solve the coupled channel equations at short-range. 

Nevertheless, the calculation can be facilitated further by using a FT approach. The general form of the FT theory as applied to ultracold collisions of two alkali atoms has been written in Refs.\cite{Burke1998, Burke1999, Gao2005}. The main simplification is to neglect the hyperfine interaction at short-range. This is justified by the fact that the exchange splitting is much larger than the hyperfine and Zeeman energy. In this case the atomic motion is described by a set of uncoupled channel equations, which can be solved numerically. Matching the solutions to the analytic base pair allows one to determine the short-range energy-analytic scattering information in terms of quantum defects $\mu^{sr}_{S}(\epsilon_{S})$ in the single-channel singlet-triplet basis (equivalently the singlet and triplet scattering lengths recast as quantum defects) in a diagonal short-range reaction matrix $K^{sr}_{\mathrm{diag}}=\mathrm{tan}(\pi \mu)$. An energy independent real orthogonal transformation turns this short-range single-channel scattering information into the final channel structure applicable at $R\rightarrow \infty$, namely the representation of hyperfine plus Zeeman atomic energy eigenstates. This procedure, delivers the real, symmetric, short-range reaction matrix $K^{sr}$ (or the corresponding smooth quantum defect matrix $\mu^{sr}$):
\begin{equation}
\label{eq:Frametransform}
K^{sr}_{ij}=\sum_{\alpha} U_{i,\alpha} \tan(\pi \mu_\alpha) \tilde{U}_{\alpha,j}.
\end{equation}
Here the tilde denotes the matrix transpose.  The dissociation channel index $i$ represents the set of quantum numbers according to Eq.~(\ref{channelsdefinition}) for non-zero magnetic field needed to characterize the internal energies of the separating atoms as well as their relevant angular momentum couplings with each other and with the orbital angular momentum quantum number $l$ and its projection $m_l=M-m_A-m_B$. 
As was stressed by Bo Gao in his ''angular momentum insensitive'' form of quantum defect theory for a van der Waals long-range potential, the $l$-dependence is known approximately \cite{Gao2001} as $\mu^{sr}_{S,l} \approx \mu^{sr}_S-l /4$~\cite{Ruzic2013}.  When higher accuracy is needed, a small $l-$dependent correction $\alpha_l$ can be introduced to this equation, which leads to: 
\begin{equation}
\label{eq:defects}
\mu^{sr}_{S,l} \approx \mu^{sr}_S-l  /4+\alpha_l,
\end{equation}
where $\alpha_0 \equiv 0$ by definition.
The FT then simply approximates the real, orthogonal matrix that diagonalizes $K^{sr}$ as the angular momentum recoupling matrix that connects the short-range eigenstates with those appropriate at large $R$.  Specifically, in the absence of any magnetic field, good quantum numbers of the atomic energy levels are given by Eq.~\eqref{channelsdefinition}. In the presence of an external magnetic field $B$ directed along the $z$-axis, $f_a$ and $f_b$ are no longer good quantum numbers but $m_{a},m_{b}$ are still conserved for the atoms at infinite separation.  However one must diagonalize the atomic hyperfine plus Zeeman Hamiltonian to obtain a numerical eigenvector $\langle f_{A} m_{A},f_{B} m_{B} | m_{A} k_{A}, m_{B} k_{B} \rangle \equiv \langle i| j \rangle $ and the corresponding field-dependent channel energies, $E_{m_{A} k_{A},m_{B} k_{B}}(B) \equiv E_{j}(B)$ (see also the extended interpretation for the basis given in Eq.~\eqref{channelsdefinition}).  We indicate the short-range collision eigenstates by $|(s_{A} s_{B})S (i_{A},i_{B})I f m_{f} \rangle \equiv | \alpha \rangle$.
We can now write out the final FT matrix $U_{i\alpha }$ between the long- and short-range channels, which is needed in Eq.~\eqref{eq:Frametransform}. Recall that in the present notation, the long-range scattering channels in the presence of a magnetic field $B\ $\ are written as $i=\{m_{A},k_{A},m_{B},k_{B}\},$ and the short-range collision eigenchannels are $\alpha =\{(s_{A},s_{B})S(i_{A},i_{B})I,fm_{f}\},$ and the unitary transformation between these is given explicitly in terms of standard angular momentum coefficients (Clebsch-Gordan and Wigner 9-j symbol) and the Breit-Rabi eigenvectors such as $\left\langle k_{A}|f_{A}\right\rangle ^{(m_{A})}$, etc. as:
\begin{equation}
\begin{split}
U_{i\alpha }= \sum\limits_{f_{A}f_{B}f}\left\langle
k_{A}|f_{A}\right\rangle ^{(m_{A})}\left\langle k_{B}|f_{B}\right\rangle^{(m_{B})} \times\\
\left\langle f_{A}m_{A},f_{B}m_{B}|fm_{f}\right\rangle \times\\
 \left\langle
(s_{A}i_{A})f_{A}(s_{B}i_{B})f_{B}|(s_{A},s_{B})S(i_{A},i_{B})I\right\rangle
^{(f)}
\end{split}
\end{equation}

Note that in the FT approximation, this matrix is
independent of $l$, so this quantum number is not explicitly represented. The transformation mentioned for the ABM is constructed in the same way.

Note that the final step of computing scattering or bound state observables such as the FRs at zero incident energy in various scattering channels requires solving the MQDT equations as a function of energy and/or magnetic field. As usual in MQDT studies, this is the step where exponential decay of the large-$R$ closed-channel radial solutions is imposed.  
The determinantal condition for a resonance to occur at an energy just above an open-channel threshold is $det(K^{sr}_{QQ}+\cot \gamma)=0,$ where the notation $K^{sr}_{QQ}$ indicates just the closed-channel partition of the full short-range $K$-matrix.  In this equation, $\gamma$ is a diagonal matrix of long-range negative energy phase parameters as mentioned above for the construction of the $S$-matrix from the MQDT.

The energy- and field-analytic nature of the single-channel solutions allows them to be constructed on a very coarse mesh of energy and magnetic field. In its most simple form, the energy dependence of the quantum defects can be dropped, and the quantum defects, which are calculated at a specific energy only once, can be used throughout the entire energy and magnetic field range of interest.  Ref. \cite{Burke1998,Gao2005} demonstrate how the quantum defects can be represented by only two parameters, namely $a_{S}$ and $a_{T}$, in the FT formalism. This yields the crudest, but also computationally lightest realization of the MQDT.

The details for a calculation of \Li-\Cs~FRs using the MQDT are given in Sect.~\ref{sec:MQDT}. Additionally, we introduce a slightly modified MQDT-FT approach, which allows us to calculate FRs for systems that are lacking a detailed microscopic model.

\section{Application to the \Li-\Cs system}
\label{sec:application}

In this chapter we apply the models described in Sect.~\ref{sec:nutshell} as a case study to the \Li-\Cs~system, where FRs have been measured recently \cite{Repp2013}. Throughout the entire section, we use the $C_{6}$ coefficient from Derevianko et al. \cite{Derevianko2001} for the description of the van der Waals interaction, which has been calculated with sufficient accuracy. In order to compare the models among themselves and with experiment, we calculate the weighted rms deviation $\delta B^{rms}$ on the resonance positions, which is defined as 

\begin{equation}
\label{eq:rms}
\delta B^{rms}=\frac{\sqrt{(\sum_{i}^{N}\delta_{i}^{2}/\delta B_{i}^{2})/N}}{\sqrt{\sum_{i}^{N}\delta B_{i}^{-2}}}. 
\end{equation}
The summation is performed over $N$ resonances, $\delta=B_{\text{res}}^{\text{exp}}  -  B^{\text{theo}}_{\text{res}}$ is the deviation of experimental ($B_{\text{res}}^{\text{exp}}$) and theoretical ($B^{\text{theo}}_{\text{res}}$) resonance positions, and $\delta B$ contains the experimental uncertainty of the measured resonance positions, which are given in Table~\ref{tab: List2}, and a $200\,\mathrm{mG}$ drift of the magnetic field for all resonances. 

\subsection{Coupled Channels Calculation}
\label{sec:channels}

We have provided details of the CC calculation for a mixture of \Li and \Cs atoms elsewhere (see Ref.\cite{Repp2013} and references therein). Here, we review the method of our CC calculation and summarize its results, as they are used as benchmark for the other approximate models presented in the subsections below.

\begin{figure}[t]
\includegraphics[width=1.0\columnwidth]{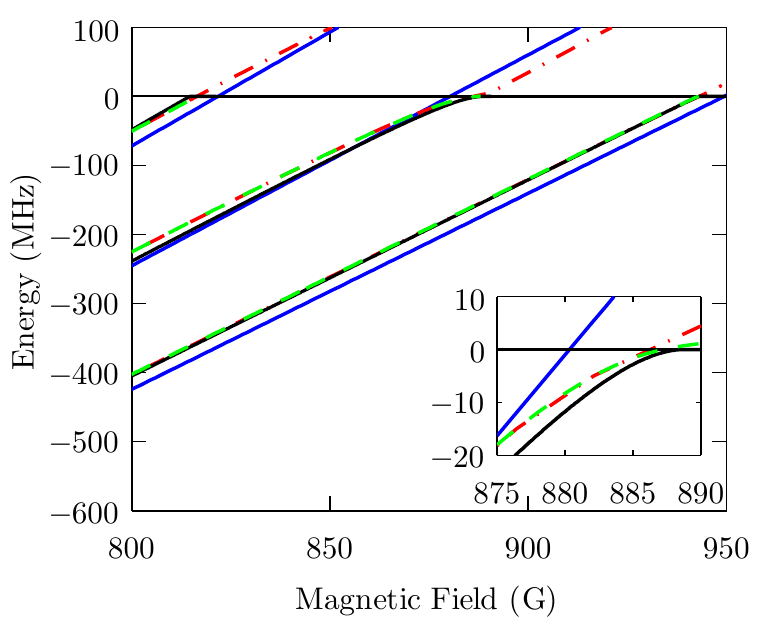}
    \caption{(Color online) Molecular energy levels for the $^6\text{Li}\left|F=1/2,m_{F}=-1/2\right\rangle\oplus  ^{133}\text{Cs}\left|3,3\right\rangle\ $ channel for $l$=0, s-waves. The energies with respect to the open channel asymptote for the bare ABM (blue line), dressed ABM (black line), MQDT (red dash-dotted line) and CC (green dashed line) are depicted. The horizontal line at zero energy represents the continuum threshold. The crossings of the molecular channels with the threshold mark the positions of the FRs. Calculated quasi-bound levels and box states from the CC model are removed for clarity. The inset shows a zoom into the region of the resonance at $\sim 889$~G, where the differences between the models is clearly visible. E.g. the energy level of the bABM is not shifted, as it neglects the coupling to the continuum.}%
   \label{fig:EnergyLevels}
\end{figure}

For the CC matrix, the Hamiltonian of Eq.~\eqref{eq:Hamiltonian} is employed, where the effective form of $H_{dd}$ (Eq.~\eqref{eq:Vdipole}) is used. Only basis states with partial waves up to $l=2$ are included, which is sufficient for the descriptions of alkali atoms in the $\mu\mathrm{K}$ regime. Besides the atomic constants, which  are readily available in the literature \cite{Arimondo1977}, accurate potentials are crucial in order to precisely determine the position of FRs. For this purpose, the relevant potential curves for the  $a ^3\Sigma^+$ and $X^1\Sigma^+$ states of \LiCs~are expanded in a power series of the internuclear separation $R$ (similar to Ref. \cite{Gerdes2008}), where $R$ is mapped onto a Fourier grid following \cite{Tiesinga1998}. Then, the expansion coefficients, which were initially determined via Fourier-transform spectroscopy \cite{Staanum2007}, are modified iteratively in such a way that both the calculated maxima of binary collision rates and the rovibrational transition frequencies are in agreement with the measured FRs and with the 6498 previously observed molecular transitions \cite{Staanum2007}, respectively.  The potential parameters are summarized in the online material, the parameters of the bound states, which are involved in the observed FRs, are given in Table~\ref{tab: List1} and the resulting resonance positions in Table~\ref{tab: List2}. The molecular energy levels for the $^6\text{Li}\left|F=1/2,m_{F}=-1/2\right\rangle\oplus  ^{133}\text{Cs}\left|3,3\right\rangle\ $ channel with respect to the incoming channel are given in Fig.~ \ref{fig:EnergyLevels}. The rms deviation for this model is 39 mG.

\begin{table*}[t]
\newcommand{\mc}[3]{
\multicolumn{#1}{#2}{#3}}
\begin{ruledtabular}
\begin{center}
\begin{tabular}{l|cc|c|cc|c|cc}
Model  & $\epsilon^{0}_{0}$ (MHz) & $\epsilon^{0}_{1}$ (MHz) & $\zeta_{0}$ & $\epsilon^{1}_{0}$ (MHz) & $\epsilon^{1}_{1}$ (MHz) & $\zeta_{1}$& $a_{S}$ ($a_{0}$) & $a_{T}$ ($a_{0}$) \\
\hline
CC  & 1566 & 3942 & 0.866  & 1159 & 3372 & 0.866 & 30.3(1) & -34.3(2) \\
bABM& 1592 & 4189 & 0.866  & 1191 & 3641 & 0.860 & \it{29.6} & \it{-42.4} \\
dABM & 1543 & 4155 & 0.870 & 1191 & 3641 & 0.860 & \it{30.7} & \it{-40.8} \\
MQDT & 1565 & 3945 & 0.866 & 1158 & 3375 & 0.862 & 30.3 & -34.4 \\
MQDT-FT & - & - & - & - & - & - & 30.1 & -39.2 \\
\end{tabular}
 \caption{List of bound state energies $\epsilon^{l}_{S}$, wave function overlaps $\zeta_{l}$ and background scattering lengths $a_{S}$ for singlet ($S=0$) and triplet ($S=1$) potentials. The fit results for the CC calculation, the bare ABM (bABM), the dressed ABM (dABM), the MQDT, and the MQDT-FT are tabulated. Note that the $l=1$ values for the bABM and the dABM are taken from the same fit. The scattering lengths indicated for the ABM are derived from the binding energies using the accumulated phase method.
}
\label{tab: List1}
\end{center}
\end{ruledtabular}

\end{table*}

\begin{table*}[t]
\newcommand{\mc}[3]{\multicolumn{#1}{#2}{#3}}

\begin{ruledtabular}
\begin{center}
\begin{tabular}{c|l|cc|c|c|c|c|l}
Entrance channel  & l& $ \text{B}_{\text{res}}^{\text{exp}} $(G)  & $\Delta \text{B}^{\text{exp}}$ (G)& $\delta_{CC}$ (G)&$\delta_{bABM}$ (G)&$\delta_{dABM}$ (G) &$\delta_{MQDT}$ (G)&   $\delta_{MQDT-FT}$ (G)\\
\hline
\Li$\left|1/2,+1/2\right\rangle \ $   & 1 & 662.79(1)& 0.10(2)&-0.04  & -0.04 & -- & -0.11& -0.26 \\
 $ \oplus \ ^{133}\text{Cs} \left|3,+3\right\rangle$  & 1 & 663.04(1)& 0.17(2) & -0.02 & -0.37 & -- & -0.01& 0.01 \\
 & 1 & 713.63(2) & 0.10(3) & -0.05  & -0.82 & -- & -0.09 & -0.22\\
 & 1 & 714.07(1) & 0.14(3) & -0.05  & -1.35 & -- & 0.10 & 0.22\\
  & 0 & 843.5(4) & 6.4(1)& 0.51& 8.32 & -0.64  & 0.38 & 0.00\\
 & 0 & 892.87(7) & 0.4(2) &-0.11  &-7.03 &  -1.20 & -0.04 & -0.39 \\
\hline
\Li$\left|1/2,-1/2\right\rangle $   & 1 & 658.21(5)& 0.2(1) &  0.07  & -3.02 & -- & 0.04 & -0.06 \\
$ \oplus \ $\Cs $\left|3,+3\right\rangle$ & 1 &708.63(1) & 0.10(2)& -0.05  & 1.24 & -- & -0.11 & -0.19 \\
 & 1 & 708.88(1) & 0.18(2)  &-0.03  & 0.91 & -- & -0.01 & 0.06 \\
 & 1 &764.23(1) & 0.07(3)  & -0.06 & 0.83 & -- & -0.06 & -0.09 \\
 & 1 &764.67(1) & 0.11(3)  &-0.05 & 0.29 & -- & 0.12 & 0.35 \\
& 0 & 816.24(2) & 0.20(4) & -0.12 & -5.19  & 1.51  & -0.26 & -0.02\\
 & 0 & 889.2(2) & 5.7(5)  &0.46  & 9.31 & 0.31  & 0.34 & 0.00 \\
 & 0 & 943.26(3) & 0.38(7)  &-0.12  & -5.68  & 0.10 & -0.04 & -0.34 \\
\hline
\Li$\left|1/2,+1/2\right\rangle \ $  & 1 & 704.49(3) & 0.35(9) & 0.07  & 0.67 & -- & 0.01 & -0.01 \\
$\oplus \ $\Cs $\left|3,+2\right\rangle$ & 0 & 896.6(7) & 10(2) & 0.68 & 19.51 & -0.35 & 0.07 & -0.21 \\
\hline
\Li$\left|1/2,-1/2\right\rangle \ $  & 1 & 750.06(6) & 0.4(2)  &0.06  &  1.47 & -- & -0.01 & 0.01 \\
$ \oplus \ $\Cs $\left|3,+2\right\rangle$ & 0  & 853.85(1) & 0.15(3) & -0.17 & -7.34 & -0.29 & -0.41 & 0.15\\
 & 0 & 943.5(1.1) & 15(3)& 2.21 & 21.69 & 1.59 & 1.64 & 1.4 \\
 \hline
$\delta B^{rms}$ (G)] & & & & 0.039 & 0.965 & 0.263 & 0.040 & 0.048\\
\end{tabular}
 \caption{Comparison of the resulting  \Li-\Cs FR positions from the various models to the observed resonances.
 The experimental positions $B_{\text{res}}^{\text{exp}}$ and widths $\Delta B^{\text{exp}}$ are taken from Ref. \cite{Repp2013}. The resonance positions $B^{\text{theo}}_{\text{res}}$ derived from CC calculation ($\delta_{CC}$), bare ABM ($\delta_{bABM}$), dressed ABM ($\delta_{dABM}$), MQDT ($\delta_{MQDT}$) and MQDT-FT ($\delta_{MQDT-FT}$) are given as deviations  $\delta$=$B_{\text{res}}^{\text{exp}}  -  B^{\text{theo}}_{\text{res}}$
with respect to the observations. We also state the rms deviation $\delta B^{rms}$ (see Eq.~\ref{eq:rms}) for all models. For the dressed ABM the $p$-wave values are identical to the bare ABM, and therefore not repeated in the table. The splitting of the $p$-wave resonances has not been considered in $\delta_{MQDT-FT}$.}

\label{tab: List2}
\end{center}
\end{ruledtabular}

\end{table*}

\subsection{Asymptotic Bound State Model}
\label{sec:ABM}

For the ABM calculation of \Li-\Cs FR positions, we begin using the ABM in its simplest form starting from the Hamiltonian of Eq.~\eqref{eq:Hamiltonian}, replacing $T+V$ by the bound-state energies and neglecting $H_{dd}$. The latter is incorporated at a later stage. Because the spacing of the vibrational states in the \LiCs~potential is large compared to the hyperfine energy, we only include the least bound vibrational state of the singlet and triplet potential and neglect the role of deeper bound states. This yields a fit of only three parameters per partial wave. However, the three fit parameters are not independent with the present set of data, as will be explained in the following discussion.

As a prelude to the new fits below, we start with the ABM as practiced in Ref. \cite{Repp2013}, where the ABM was applied leaving five parameters ($\epsilon_0^0$, $\epsilon_1^0$, $\zeta_0$, $\epsilon_0^1$, $\epsilon_1^1$) as free fit parameters, while $\zeta_1$ was taken to be equal to $\zeta_0$ \cite{[{Also, the atomic masses in the calculation where not accurate enough, which also had minor effects on the fitting results. In the present work we use atomic masses from~}][{ (version 3.0). [Online] Available: http://physics.nist.gov/Comp [2013, 12 09]. National Institute of Standards and Technology, Gaithersburg, MD.}]atomicmasses2012}. 
In this work we redo the fit, utilizing $\zeta_{1}$ also as a free parameter, thus using six parameters as fit parameters to minimize $\delta B^{rms}$ (see Eq.~\eqref{eq:rms}). This quantity gives intuitive and quantitative insight into the deviations of calculated from measured resonance positions. In \Li-\Cs the hyperfine interaction gives rise to a very strong singlet-triplet mixing, which is indicated by an expectation value for the total spin $S$ of $\left\langle S \right\rangle \simeq 0.6-0.7$ on the resonances. This results in a strong correlation of the fit parameters and the resonances can be fitted with similar rms deviations over a large range (within a few GHz) of binding energies.
The best fit has a rms deviation of 877~mG with the parameters $\epsilon_0^0=5824$~MHz, $\epsilon_1^0=2995$~MHz, $\zeta_0=0.559$, $\epsilon_0^1=1844$~MHz, $\epsilon_1^1=3575$~MHZ and $\zeta_1=0.821$. However, since the singlet and triplet binding energies and their overlap parameter are related to each other through the interaction potential, the obtained overlap parameter of $\zeta_{0}=0.559$ is unphysical for the fitted binding energies. Additionally, the $p$-wave shift is unreasonably large for the singlet channel binding energy, while it has opposite sign for the triplet binding energy, which are clear indications that the fit results are unphysical.

In order to obtain a fit restrained to physical parameters, we demonstrate how these discrepancies of binding energies and overlap parameters can be reduced in the bare ABM (bABM) in three steps. The first step is to reduce the number of independent fit parameters by deriving the wave function overlaps from the two binding energies via the accumulated phase method (as described in Ref.~\cite{Verhaar2009}) instead of leaving them as a free parameter, thereby restricting ourselves to only the physical range of the fit parameters and reducing the fit to two parameters.    
The coupling of the bound state to the continuum, which is neglected at this stage, results in significant shifts for $s$-wave resonances. Therefore, we use only the narrow $p$-wave resonances for the initial fit, because their widths are not acquired by coupling to the continuum but rather by tunneling through the centrifugal barrier which is suppressed at low collision energies. 
The fit results of $\epsilon^{1}_{0}=1193 \  \text{MHz}$, $\epsilon^{1}_{1}=3638 \  \text{MHz}$ and the calculated $\zeta_{1}=0.861$
agree much better with the CC values, which are shown in Table~\ref{tab: List1}. 
The $p$-wave resonances are reproduced with a rms deviation  of 560 mG, where the mean was used for resonances which are split due to the magnetic spin-spin and second-order spin-orbit coupling. This demonstrates how the bare ABM model, which extensively simplifies the spatial part of the scattering problem, satisfactorily reproduces resonances which are not shifted due to coupling to the open channel scattering wave function.

Since we obtain the asymptotic wave functions in the procedure described above, in the second step it is now also possible to include the  magnetic dipole-dipole and second-order spin-orbit coupling term  into the ABM Hamiltonian (similar to Ref. \cite{Goosen2010}) in its effective form (see Eq.~\eqref{eq:Vdipole}).  With a rms deviation of $375$\,mG, a fit containing $H_{dd}$ improves the prediction of the $p$-wave resonances by about $185$\,mG, which is the expected order of magnitude, considering that the splitting is only on the order of a few hundred mG at most. For calculating the expectation value of $H_{dd}$ only the long range part of the wave function was used. Thus the spin-orbit contribution does not play a role and  $a_\mathrm{SO}$ could be set to zero.


In order to include the $s$-wave resonances in the third and final step of the bABM, 
the $s$-wave binding energies are deduced from the fitted $p$-wave binding energies using the accumulated phase method as follows. We numerically solve the Schr\"odinger equation containing only the van der Waals term in the interaction potential and using the phases of the obtained $p$-wave functions at $R_{i}$ and $\psi \rightarrow 0$ for $R\rightarrow \infty$ as boundary conditions. Here, $R_{i}$ is the radius where the van der Waals energy is larger than the hyperfine energy, and the exchange energy is large enough to split the singlet-triplet manifold~\cite{Verhaar2009}. 
This approach neglects the $l$-dependence of the phase-shift at $R_{i}$ which is a small correction in our case \cite{Verhaar2009}. 
The rms deviation for all resonances in the case where the $p$-wave resonances are fitted and the $s$-wave resonances are calculated is 1.26 G when $H_{dd}$ is neglected. 
Including $H_{dd}$ into this fit results in the lowest attainable rms deviation value of 965 mG for a physically meaningful bare ABM fit. The resulting molecular energy levels for the $^6\text{Li}\left|F=1/2,m_{F}=-1/2\right\rangle\oplus  ^{133}\text{Cs}\left|3,3\right\rangle\ $ are shown in Fig.~\ref{fig:EnergyLevels}. As one can see, the positions where the bound state energies cross the threshold are not shifted due to interactions, as is the case for the other models. Thus, compared to the measured values, the broad FRs are systematically shifted to lower magnetics fields, as illustrated in Table~\ref{tab: List2}, where the resonance positions are given in the "bABM" column. The positions of the narrow resonances are shifted to higher values most likely  because of the application of the accumulated phase methods, which introduces errors in the determination of the $s$-wave binding energies. The latter are given together with the $p$-wave binding energies as "bABM" values in Table~\ref{tab: List1}. If one needs a more accurate prediction for the narrow resonances, one could also fit them using $\epsilon^{0}_{0}$ and $\epsilon^{0}_{1}$ as additional parameters. 

The binding energies can be used to derive background scattering lengths via the accumulated phase method \cite{Verhaar2009}, by propagating the wavefunction with the known phase at $R_{i}$ to large internuclear separations and then comparing to a long-range wavefunction which is not shifted by an interaction potential. This procedure introduces additional errors on the order of $\sim 10 \%$ that depend somewhat on the precise choice of $R_{i}$. These errors are related to the accumulated phase method and not to the ABM. By including the energy dependence of the accumulated phase as described in Ref.~\cite{Verhaar2009} the scattering length might be calculated with better accuracy. However, calculating these derived quantities allows for a comparison with the MQDT-FT (see Sect.~\ref{sec:MQDT})  where no binding energies were derived directly, and yields an additional test of consistency among the three different models. The results are shown in Table~\ref{tab: List1}.

In order to achieve higher accuracy, we include the coupling of the closed channel responsible for the FR to the open channel, which is referred to as dressed ABM (dABM). In Sect. \ref{sec:NutshellABM} we discussed the limiting cases, where the coupling of the closed channel bound state to either the least bound state, or to a virtual state in the open channel can be used as an estimate for the shift of the resonance position. The \Li-\Cs system, however, is in an intermediate regime, where both of the approaches do not deliver satisfactory results. The background scattering length of the triplet channel of $a_{T}=-34.3(2) \ a_{0}$ indicates this regime. On the one hand, it is significantly far from the van der Waals range of $r_{0}=45 \ a_{0}$, which would indicate a non-resonant open channel, but on the other hand, it is also not dominating the scattering process, which would lead to a much larger magnitude of the background scattering length. 

Therefore, we introduce an extension to the ABM similar to the approach presented in Park, et al. \cite{Park2012}, which includes both effects in a phenomenological manner. We use the complex energy shift of Eq.~\eqref{eq:VScomplexShift}, just as in a system with a resonant open channel \cite{Marcelis2004,Kempen2004}. To calculate $A_{vs}=A\zeta_{vs}$, we multiply the square of the appropriate matrix element of $\mathcal{H}_{PQ}$ taken from the matrix $\underline{M'}_{ABM}$ in the form where $\mathcal{H}_{QQ}$ is diagonalized (denoted by $\mathcal{K}^2$) with an additional scaling factor $\zeta_{vs}$, which handles the spatial part of the matrix element and is equal for all FRs. However, in contrast to Ref. \cite{Marcelis2004}, the molecular energy $\epsilon_{Q}$ is not taken to be the bare energy from the submatrix $\mathcal{H}_{QQ}$, but we rather diagonalize the full matrix  $\underline{M}_{ABM}$ and replace $\epsilon_{Q}$ in the $S$-matrix with the dressed resonant molecular state $\epsilon_{ABM}$. In doing so, $\epsilon_{ABM}$ contains the coupling to the open channel bound state. Therefore, both the influence of the coupling to the bound and the virtual states in the $P$-channel are accounted for. The FR positions are now simply obtained at magnetic fields where the following identity is satisfied:
\begin{equation}
\epsilon_\mathrm{ABM} = \frac{\mu}{\hbar^2}\frac{\zeta_{vs} \mathcal{K}^2}{\kappa_{vs}^2}.
\label{eq:New_ABM_energy}
\end{equation}
$\kappa_{vs}$ is obtained as described in Sect.~\ref{sec:NutshellABM} and $\zeta_{vs}$ is left as a free fit parameter. Within this approach the FRs, including coupling to the (near-resonant) scattering states, can be found by simple matrix operations and linear equations.

We note that the full scattering properties around the resonance (including the resonance width and for the case of overlapping resonances \cite{Park2012}), can be obtained from the $S$-matrix by using the complex energy shift $A_{vs}$ as given above.

We start the calculation by only fitting the narrow $s$-wave resonances, where the coupling to the open channel is small, in order to derive $\epsilon^{0}_{0}$ and $\epsilon^{0}_{1}$. Their overlap $\zeta_{0}$ is obtained with the same method as for the bare ABM. The results of this fit are given in Table~\ref{tab: List1} as "dABM", where the background scattering lengths for the specific incoming channels are deduced in the same manner as described above and are used to calculate $\kappa_{vs}$ for each channel needed for Eq.~\eqref{eq:New_ABM_energy}. The next step is the fitting of the scaling factor $\zeta_{vs}$ by performing a weighted least squares minimization on all observed $s$-wave resonances, additionally allowing the overlap parameter to vary by $<0.1\%$.  The result of $\zeta_{vs}=0.0255$ yields a rms deviation of 310 mG on all $s$-wave resonances.
Combining with the results of the bare ABM for the $p$-wave resonances we obtain a total rms deviation of $\delta B^{rms} = 263\,\mathrm{mG}$ on all resonances. The resonance positions are listed in Table~\ref{tab: List2}, and selected molecular energy levels are given as a function of magnetic field in Fig.~\ref{fig:EnergyLevels} for comparison. The largest deviation between the CC and ABM approach is seen at the broad resonances at 890~G (see inset Fig.~\ref{fig:EnergyLevels}). The coupling to the continuum is obvious by the nonlinear function of the energy with respect to the magnetic field. This is in contrast to the bare ABM, where the influence of the continuum is neglected.

While the ABM assigns the FRs with a sub-1 G accuracy, there is a significant deviation of the triplet binding energies $\epsilon_1^0$ from the CC value. This could arise from the fact that the coupling to the (near resonant) triplet channel is only included phenomenologically by adding a single virtual state. Also, only one bound state is taken into account and we treat the resonances as being non-overlapping.

An additional approximation is introduced by using the accumulated phase method to derive the overlap parameters, and relating the $s$-wave and $p$-wave binding energies enlarges the uncertainty. We characterize the accuracy of the accumulated phase method by comparing the obtained binding energies with the bound states of the full \Li-\Cs~potentials. Using the boundary conditions at $R_i$ from the $s$-wave binding energies we find that the accumulated phase method reproduces the $p$-wave binding energies to $\sim 1-2\%$ for both the singlet and triplet potentials. Changing the binding energies on the order of $\sim 1-2\%$ increases $\delta B^{rms}$ on the FRs with more than a factor 2, indicating that at this level of accuracy the inner part of the potential has a significant effect. Extending the ABM to use more information of the full potentials to link the $s$- and $p$-wave binding energies is straightforward, however, by this step we would lose the advantage of the simple calculations of the ABM. We also note that a more rigorous method to include both the bound and the virtual state is the Resonant State Model as presented in Ref. \cite{Goosen2011}. Also, the accumulated phase methods leads to an error in the derivation of the background scattering lengths. While this is on the order of $\sim 10 \%$, a significant part of the deviation of $a_{T}$ is a result of the systematic shift of $\epsilon^{0}_{1}$, in both the bare and the dressed ABM.

\subsection{Multichannel Quantum Defect Theory}
\label{sec:MQDT}

We apply the {\it ab initio} MQDT treatment, using the potentials of Ref.~\cite{Repp2013} as input for the calculation. We then slightly modify the inner wall of these potentials in order to minimize the deviation from the experiment. The present calculation only requires solving the coupled differential equations out to $r_0=$40 a.u., and very little difference is seen if this matching radius to the long-range single-channel QDT solutions is reduced to $30$ a.u. Table~\ref{tab: List2} shows the accuracy of the FR positions in comparison with the experimentally determined resonances of Repp et al.\cite{Repp2013}, and Fig.~\ref{fig:EnergyLevels} plots three of the obtained energy levels for comparison with the other models. The close agreement between CC and MQDT is very satisfying. The bound state energies (see Table~\ref{tab: List1}), which do not play the same central role in the MQDT calculation as in the ABM model, can be extracted from the underlying modified potentials for comparison. They show excellent agreement with the CC values and give a measure as to what degree the potentials from the CC calculation have been modified. This agreement and the small rms deviation of the FR positions from the experimental values, as given in Table~\ref{tab: List2} demonstrate, that MQDT and CC calculation are asymptotically consistent with regard to $\delta B^{rms}$ but for the description of individual resonances the two models deviate up to $\sim 100$~mG. This might also indicate the limit for predicting new FRs.

In the present study we test an alternative way to utilize the FT plus MQDT formulation; the idea is to empirically fit the single-channel singlet and triplet quantum defects so as to achieve optimum agreement with a few measured FRs.  In this treatment, if the long-range  van der Waals coefficient is already known to sufficient accuracy, as is believed to be the case for \Li-\Cs, then with two fit parameters it is possible to achieve good agreement with all of the $s$-wave resonances that have been measured to date, and to predict additional resonances.  The $l$-dependence of the fitted quantum defects is approximately known, but to achieve better accuracy on other partial waves, it appears to be necessary to fit one small additional correction for $p$-waves (see Eq.~\eqref{eq:defects}).  While the MQDT has been shown in a number of studies to give a highly efficient way to calculate ultracold scattering observables when the interaction potentials are known, there is an increasing demand for a robust method for analyzing new, complex systems where FRs have been measured but not yet analyzed to the level of yielding a detailed microscopic model.  The present test of the semi-empirical MQDT frame transformation (MQDT-FT) is encouraging in its potential for such problems, as is seen from the results presented below for the \Li-\Cs interaction.

In our implementation of the frame-transformed version of MQDT utilized for the present study, the long-range parameters ($A$, $\mathcal{G}$, $\gamma$ and $\eta$) are determined once and for all for a pure van der Waals potential at long-range, $-C_6/R^6$. The long-range MQDT parameters are standard and can be used for any alkali atom collision, because they are tabulated as functions of the single dimensionless variable which is the product of the van der Waals length and the wavenumber $k$ (see e.g. Ref.~\cite{Ruzic2013}). Two energy-independent and field-independent short-range quantum defects, namely $\mu^{sr}_{S,l}$ for $S=0,1$, were adjusted until optimum agreement was achieved with the experimental resonance positions. Note that a global search was not carried out over all values of the $0 \le \mu^{sr}_S < 1$ (mod 1).  The starting values of the search came from quantum defects extracted from the \Li-\Cs~potentials provided in Ref. \cite{Repp2013} and only small adjustments of those values were needed in the MQDT-FT fit to achieve the quoted level of agreement with the experimental resonance locations. In contrast with the MQDT calculation and the full CC calculation, this MQDT-FT calculation did not include the magnetic dipole-dipole interaction nor the second-order spin-orbit interaction term.  Nevertheless, the fit with three adjustable parameters (adding $\alpha_1$ of Eq.~\eqref{eq:defects}) gives a small rms deviation with experimental resonance positions in  Table~\ref{tab: List2}, namely 47 mG. The fitted $p$-wave correction is $\alpha_1=0.00208$ and the $l=0$ quantum defects are $ \left \lbrace \mu^{sr}_{0},\mu^{sr}_{1} \right\rbrace=\left\lbrace   0.092115, 0.346848 \right\rbrace $. These can be used to derive the background scattering lengths of the singlet and triplet potential, which are given in Table~\ref{tab: List1}. Because the bound state energies are only expected to be accurate when the binding is quite small, no comparisons with bound levels obtained in the other methods are presented here. In Table~\ref{tab: List1} one sees that the largest deviations of the scattering lengths appear for the triplet state. Despite the very different qualities of the fit for the ABM and MQDT-FT approach, their results for the triplet scattering lengths are fairly close but deviate significantly from the result of CC and MQDT. Similarly, the fit quality of MQDT and MQDT-FT are comparable but the derived scattering lengths deviate strongly. No physical reason for this behavior is known at present.


\section{Conclusion}

\label{sec:Conclusion}

In this work we have applied three different models for the assignment of the measured FRs. All three models describe the observed resonances with a sub-1 G accuracy. However, depending on the desired degree of precision and the availability of accurate interaction potentials, the models serve different purposes.

In some cases a phenomenon under investigation requires highly accurate knowledge of scattering observables that are not measured but rather deduced from theory, as for example the scattering length in dependence of the magnetic field. In these cases it is inevitable to use the CC calculation. This demands either accurate  \textit{ab initio} potentials or sufficient experimental data to construct such potentials. The high accuracy of the CC calculation stems from the fact that it incorporates the least amount of assumptions out of the three models. The rms deviation of 39 mG is the lowest of all three models. It is a rigorous and straightforward numerical approach, which comes at the cost of computational power and complex codes. The potentials for the CC calculations were mainly determined by spectroscopic data and here the data set for the triplet state is fairly sparse. The minimum is not well characterized because the vibrational levels from $v=0-4$ are not yet observed. Thus, predictions of FR for $^7$Li-\Cs may be less accurate than for \Li-\Cs. Measurements would be very desirable.

There are situations where the required precision of scattering observables is less stringent. One example is evaporative cooling of ultracold gases or sympathetic cooling of ultracold mixtures to quantum degeneracy by means of a FR, or for initial characterization of FRs. In these cases, the processes are often optimized experimentally, and it is not necessary to know the exact value of the scattering length for the start of the optimization. Under these circumstances the two other simplified models are much more appropriate. 

With a rms deviation of 965 mG, the bare ABM explains the FR structure already on the level of $\sim$ 1 G. Only two parameters are sufficient for the description of the FR positions. The relatively large deviation is related to the fact that couplings to continuum states are neglected, which results in shifts on the order of the FR width of broad resonances. The fact that the rms deviation of $p$-wave resonances is only 560 mG, and can be further reduced to 375 mG when the spin-spin interaction is included, shows that narrow resonances are predicted sufficiently well. An advantage of the bare ABM is that no molecular information is required, as it builds solely on atomic constants and few fit parameters. Additionally, the code for the calculation is extremely simple since it only involves the numerical diagonalization of a small matrix, which is included in standard computational software programs. Therefore, it can be applied at low programming expense for all systems, in order to assign or predict FRs, or quickly map out all resonances of a system. In fact, it was used to estimate whether there are any FRs expected in experimentally achievable field regions for the \Li-\Cs ~system before the experiment was set up. Also, it can be used to optimize the starting conditions for a CC calculation \cite{Li2008}. We remark that the open channels for $M=5/2$ and $M=3/2$ have overlapping continua, $\left|1/2,-1/2\right\rangle  \oplus \ $ $\left|3,+3\right\rangle$ with  $\left|1/2,1/2\right\rangle  \oplus \ $ $\left|3,+2\right\rangle$ or $\left|1/2,-1/2\right\rangle  \oplus \ $ $\left|3,+2\right\rangle$ with $\left|3/2,-3/2\right\rangle  \oplus \ $ $\left|3,+3\right\rangle$, respectively. Thus the two-state approach for the $S$-matrix might not be completely justified.

For the calculation of scattering properties and an accurate description of broad resonances, the dressed ABM has to be applied. With a rms deviation of 263 mG it is somewhat less accurate than the MQDT-FT. Yet, since it does not solve the Schr\"odinger equation numerically but rather utilizes an analytical expression of the $S$-matrix, it is still computationally straightforward. In cases where the background scattering lengths for the incoming channel are known, the implementation is simple, since a comparison with the van der Waals length directly shows whether the analytical expressions for bound or virtual state or --as is the case for \Li-\Cs-- for an intermediate regime are appropriate. If this information is not available, we suggest first fitting the narrow resonances, in order to deduce background scattering lengths from the fitted bound state energies via the accumulated phase method, as explained in Sect.~\ref{sec:ABM}. Then the right choice of the analytical expression to be used will become evident.

In order to gain accurate information from the MQDT, the interaction potentials have to be known sufficiently well. However, compared to the CC calculation, it reduces the complexity of the problem enormously. As it is still a full scattering physics approach, employing a coupled-channel solution at short-range, the code is more complex and lengthy than the ABM code. Yet, once this code is available, it can be used for any alkali system without adaptation to the system specifics and solves the scattering problem efficiently. The accuracy of the final results with a rms deviation of 40 mG is statistically indistinguishable from that of the full CC calculation. Also, both yield smaller values for the FR positions of the broad resonances as compared to the observations. This deviation stems from the fact that for the experimental determination of the position, a Gaussian profile was fit to the loss spectrum, which neglects the asymmetric line shape of a broad resonance and therefore returns slightly larger values than the actual resonance positions. As a result, the current investigation does not indicate a model problem of CC and MQDT for the broad resonances.

Many of the above mentioned properties of the MQDT are also true for the MQDT-FT. The latter is especially useful for systems with little knowledge of interaction potentials and only a few experimentally measured FRs. A two parameter fit for only $s$-waves (three-parameters for $s+p$ etc.) allows to assign the resonances and to investigate the existence of possibly broader or for specific applications more appropriate FRs. While the rms deviation of 48 mG is comparable to MQDT and CC models, the predicted values of the bare singlet and triplet scattering lengths is less accurate. Because the variation of the quantum defects compensates for deviations introduced by the assumptions of the MQDT-FT (see Sect.~\ref{sec:MQDT}) in order to recreate the FR positions, the accuracy of other scattering properties should be tested in future studies. As it relies on only three parameters for the prediction or assignment of FR, it is appropriate in systems that are currently lacking accurate interaction potentials.

In conclusion, depending on the knowledge of molecular parameters, the required accuracy of the predicted scattering parameters, the complexity of code and the computational expense, each model has its own strength in applicability.

\section{Acknowledgments}
The work carried out at Colorado and at Purdue has been supported by the U.S. Department of Energy, Office of Science. The work carried out in Heidelberg was supported by the CQD. R.P. acknowledges support by the IMPRS-QD. J.U. acknowledges support by the DAAD.
\section{Appendix}

\begin{figure}[h]
\subfigure[$\Delta\approx-0.167$ G, and $a_{1,\text{bg}}^3(B)$ is constant.]{
\includegraphics[width=.99\columnwidth]{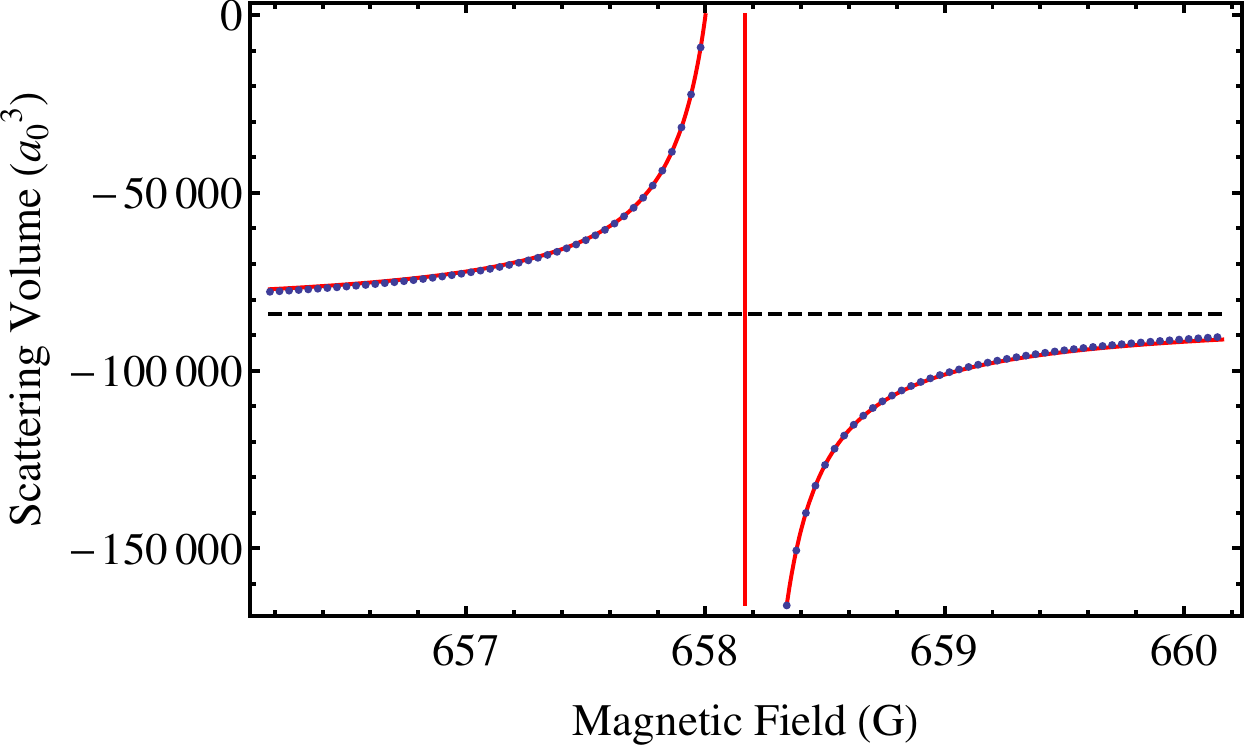}
\label{fig:polea}
}
\subfigure[$\Delta\approx-1.65$ $\mu$G, and $a_{1,\text{bg}}^3(B)$ is linear.]{
\includegraphics[width=.99\columnwidth]{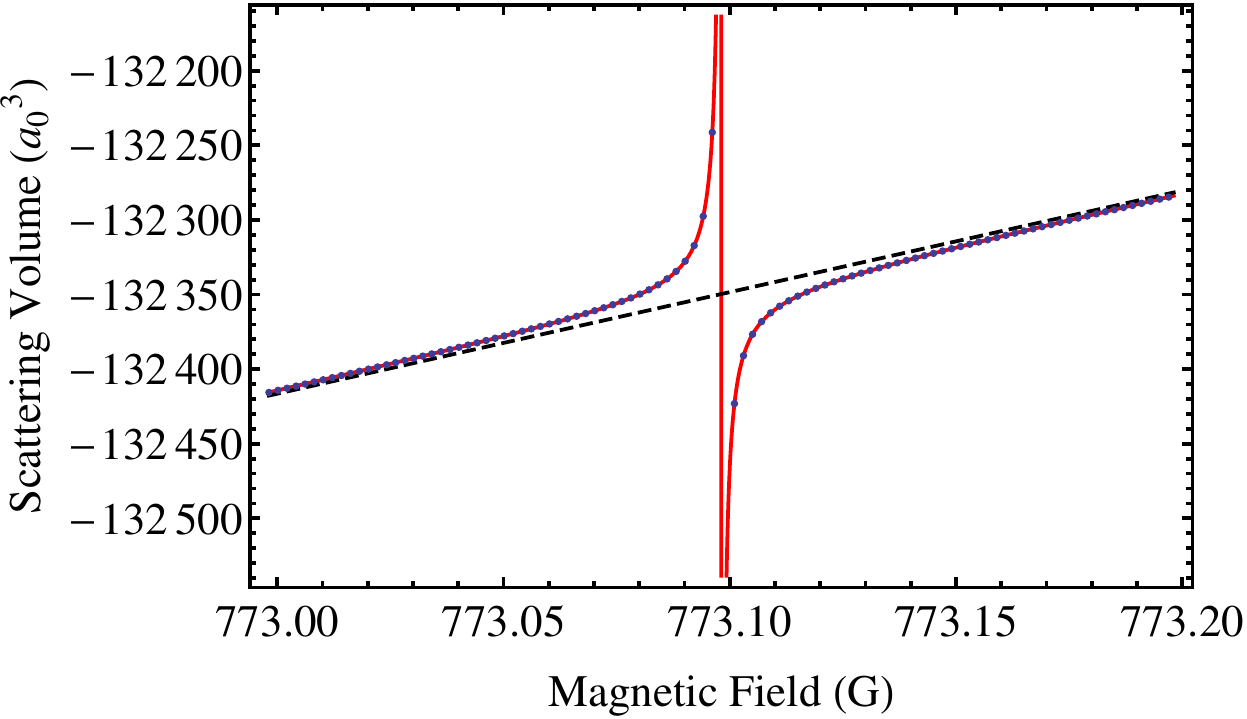}
\label{fig:poleb}
}
\caption{(Color online) These graphs show the $l=1$ resonances near (a) 658.2 G and (b) 773.1 G. The (red) curves are the best fit form of Eq.\eqref{eq:pole}. The (blue) dots are the MQDT calculation of $a_1^3(B)$. The (black) dashed line is the function $a_{1,\text{bg}}^3(B)$.}
\end{figure}

In this Appendix we demonstrate the predictive power of the MQDT calculation by determining all the $s$- and $p$-wave FRs for the initial states measured experimentally, in the magnetic field range 0-1500 Gauss. For brevity we only describe $p$-wave FRs for a single value of incident $m_l$ for each incident spin state. We also describe how the theory extracts resonance widths.  

Within MQDT finding and identifying FRs is straightforward. Approximate FR locations are quickly determined by searching for roots of det($K^\text{sr}_{QQ}+\cot\gamma$), and the eigenstate of $K^\text{sr}_{QQ}+\cot\gamma$ whose eigenvalue crosses zero near an FR identifies the quantum numbers of the resonant state. Table \ref{tab:key} reports FRs in the range of magnetic field $B=0-1500$ G, labeled by their incident spin state, $\ket{f_{\text{Li}}, m_{f_{\text{Li}}}, f_{\text{Cs}}, m_{f_{\text{Cs}}}}$, and resonant-state quantum numbers, $m_{f_{\text{Li}}}+m_{f_\text{Cs}}$, $l$, and $m_l$. The $m_l$ quantum number of each incident channel is easily inferred from Table \ref{tab:key} by conservation of total angular momentum, $m_\text{tot}^\text{inc}=m_\text{tot}^\text{res}$.

To characterize each FR in terms of a position $B_0$ and a width $\Delta$ in magnetic field, we calculate one of two quantities: the real part of the scattering length, $a_0^1$, or the real part of the scattering volume, $a_1^3$. We refer to these two quantities simultaneously as $a_l^{2l+1}$. MQDT quickly generates $a_l^{2l+1}$ on a fine grid in magnetic field, and we fit $a_l^{2l+1}(B)$ to one of three different functional forms described below.

For the majority of FRs in Table \ref{tab:key}, the resonant state is much more strongly coupled to the incident scattering channel than to any inelastic (exoergic) channel, and a clear pole emerges in $a_l^{2l+1}(B)$. In this case $a_l^{2l+1}(B)$ takes the conventional form,
\begin{equation}
\label{eq:pole}
 a_l^{2l+1}(B)=a_{l,\text{bg}}^{2l+1}(B)\left(1-\frac{\Delta}{B-B_0}\right),
\end{equation}
where $\Delta$ and $B_0$ are constants. $\Delta$ is the field width, and $B_0$ is the resonance location.

When $|\Delta|$ is relatively large ($|\Delta|>0.1$ G), we let $a_{l,\text{bg}}^{2l+1}(B)$ be constant in $B$. This allows for an excellent fit of $a_l^{2l+1}(B)$. However, when $|\Delta|$ is relatively small ($|\Delta|<0.1$ G), we let $a_{l,\text{bg}}^{2l+1}(B)$ be linear in $B$ to achieve an equivalent fit. Figures \ref{fig:polea} and \ref{fig:poleb} demonstrate fits of resonances described by Eq.~\ref{eq:pole} when $|\Delta|>0.1$ and $|\Delta|<0.1$, respectively.

For several FRs in Table \ref{tab:key}, the resonant state is comparably coupled to both the incident channel and an inelastic channel, and the variation of $a_l^{2l+1}(B)$ becomes less drastic than for pure elastic scattering. In this case we fit $a_l^{2l+1}(B)$ to the form~\cite{Hutson2007},
\begin{equation}
\label{eq:inelastic}
 a_l^{2l+1}(B)= a_{l,\text{bg}}^{2l+1}(B) + \frac{\alpha \big(2(B-B_0)/\Gamma\big) + \beta}{\big(2(B-B_0)/\Gamma\big)^2+1},
\end{equation}
where $\alpha$, $\beta$, $\Gamma$, and $B_0$ are constants. $\Gamma$ is the inelastic field width.

\begin{figure}[h]
\includegraphics[width=.99\columnwidth]{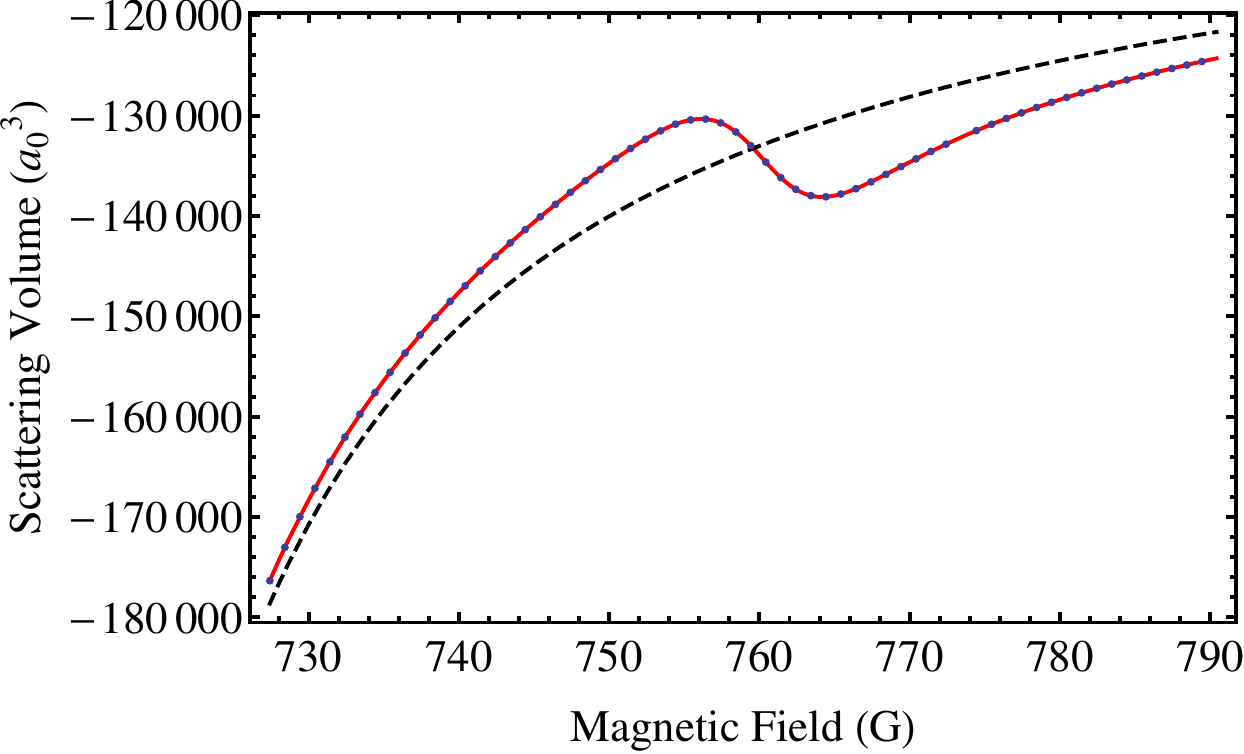}
\caption{\label{fig:inelastic} (Color online) This graph shows the $l=1$ resonance near 760.4 G. The (red) curve is the best fit $a_1^3(B)$ from Eq.~\ref{eq:inelastic}. The (blue) dots are the MQDT calculation of $a_1^3(B)$. The (black) dashed line is the high order polynomial $a_{1,\text{bg}}^3(B)$. This fit excludes data around the narrow resonance near 773.1 G.} 
\end{figure}

Since the variation in $a_l^{2l+1}(B)$ near these FRs can be very small, we fit these FRs by letting $a_{l,\text{bg}}^{2l+1}(B)$ be a high order (order=9) polynomial in $B$. This high order fit is appropriate as the coefficients decrease by orders of magnitude with successive powers of $B$, and the best fit $a_{l,\text{bg}}^{2l+1}(B)$ is not oscillatory in the vicinity of $B_0$. For example, Figure \ref{fig:inelastic} shows the fit of the resonance near 760.4 G.

We calculate all FRs at 1 $\mu$K incident collision energy. $a_l^{2l+1}$ is approximately independent of energy on this ultralow energy scale,
\begin{equation}
 a_l^{2l+1}\xrightarrow{k\rightarrow0}-\tan\delta_l/k^{2l+1},
\end{equation}
where the phase shift, $\delta_l$, obeys the Wigner threshold laws, $\delta_l\xrightarrow{k\rightarrow0}\propto k^{2l+1}$. We numerically determine that the simple threshold behavior of $\delta_l$ leads to energy independent field widths, $\Delta$ and $\Gamma$, for $l=0$ and $l=1$ resonances.

Table \ref{tab:key} summarizes the behavior of each FR in terms of the small set of parameters: $B_0$, $\Delta$, $a_{l,\text{bg}}^{2l+1}(B_0)$, and $\Gamma$. We use the parameter $\Delta$ in order to directly compare all FRs, regardless of their character. As fitting the FRs to the form of Eq.~\ref{eq:inelastic} does not determine $\Delta$, we suggest an approximate relation between $\Delta$ and $\Gamma$. By comparing the large $(B-B_0)/\Gamma$ limit of Eq.~\ref{eq:inelastic} to Eq.~\ref{eq:pole}, we obtain,
\begin{equation}
  \Delta\approx -\frac{\alpha \Gamma/2}{a_{l,\text{bg}}^{2l+1}(B_0)}.
\end{equation}
Listing a value for $\Gamma$ in Table \ref{tab:key} indicates this approximation for $\Delta$.

Five of the resonances whose locations are identified by MQDT do not exhibit an appreciable variation in $a_{l}^{2l+1}$ with $B$. For these FRs Table \ref{tab:key} gives the predicted resonance location from the root of det($K^\text{sr}_{QQ}+\cot\gamma$) but does not report a width $\Delta$. Our method has found FRs with a $\Delta$ as small as $10^{-9}$ G; therefore, the uncharacterized resonances are most likely heavily suppressed by inelastic scattering or extremely narrow.

\begin{widetext}
\begin{table*}[t!]
\caption{\label{tab:key} This table characterizes FRs in $^6$Li + $^{133}$Cs obtained from the MQDT by reporting the incident spin state, $\ket{f_{\text{Li}}, m_{f_{\text{Li}}}, f_{\text{Cs}}, m_{f_{\text{Cs}}}}$; resonant state quantum numbers, $m_{f_{\text{Li}}}+m_{f_\text{Cs}}$, $l$, and $m_l$; resonance location, $B_0$; field width, $\Delta$; background value of $a_l^{2l+1}(B)$, $a_{l,\text{bg}}^{2l+1}(B_0)$ (in atomic units); and inelastic field width, $\Gamma$. The incident collision energy in each case is 1 $\mu$K. All magnetic field values are in units of gauss. When $l=0$ the background scattering length, $a_{0,\text{bg}}(B_0)$, has units of $a_0$. When $l=1$ the background scattering volume, $a_{1,\text{bg}}^3(B_0)$, has units of $a_0^3$.
}
\begin{ruledtabular}
\begin{tabular}{cccccccc}
$\ket{f_{\text{Li}}, m_{f_{\text{Li}}}, f_{\text{Cs}}, m_{f_{\text{Cs}}}}$ &$m_{f_{\text{Li}}}+m_{f_\text{Cs}}$ & $l$ & $m_l$ & $B_0$ & $\Delta$ & $a_{l,\text{bg}}^{2l+1}(B_0)$ & $\Gamma$ \\
\hline
$\ket{1/2,	1/2,	3,	3}$	&	5/2	&	1	&	1	&	   634.2	& $	 -1.39 \times 10^{-4}	$ & $	 -6.89 \times 10^{4}	$ & $	-	 $ \\
$	$	&	7/2	&	1	&	0	&	   662.9	& $	 -9.55 \times 10^{0}	$ & $	 -1.02 \times 10^{5}	$ & $	-	 $ \\
$	$	&	5/2	&	1	&	1	&	   682.3	& $	 -3.98 \times 10^{-6}	$ & $	 -1.52 \times 10^{5}	$ & $	-	 $ \\
$	$	&	9/2	&	1	&	-1	&	   690.6	& $	 -2.50 \times 10^{-5}	$ & $	 -1.36 \times 10^{5}	$ & $	-	 $ \\
$	$	&	7/2	&	1	&	0	&	   713.7	& $	 -5.92 \times 10^{-1}	$ & $	 -1.23 \times 10^{5}	$ & $	-	 $ \\
$	$	&	5/2	&	1	&	1	&	   737.6	& $	 -2.04 \times 10^{-9}	$ & $	 -1.20 \times 10^{5}	$ & $	-	 $ \\
$	$	&	7/2	&	0	&	0	&	   843.1	& $	 -6.56 \times 10^{1}	$ & $	 -2.64 \times 10^{1}	$ & $	-	 $ \\
$	$	&	7/2	&	0	&	0	&	   892.9	& $	 -2.07 \times 10^{0}	$ & $	 -6.40 \times 10^{1}	$ & $	-	 $ \\
$\ket{1/2,	-1/2,	3,	3}$	&	3/2	&	1	&	1	&	   632.5	& $	 -2.01 \times 10^{-6}	$ & $	 -9.01 \times 10^{4}	$ & $	-	 $ \\
$	$	&	5/2	&	1	&	0	&	   658.2	& $	 -1.67 \times 10^{-1}	$ & $	 -8.42 \times 10^{4}	$ & $	-	 $ \\
$	$	&	3/2	&	1	&	1	&	   676.0	& $	 -9.57 \times 10^{-5}	$ & $	 -7.46 \times 10^{4}	$ & $	-	 $ \\
$	$	&	7/2	&	1	&	-1	&	   687.4	& $	-	$ & $	-	$ & $	-	 $ \\
$	$	&	5/2	&	1	&	0	&	   708.7	& $	 -9.32 \times 10^{0}	$ & $	 -1.03 \times 10^{5}	$ & $	-	 $ \\
$	$	&	3/2	&	1	&	1	&	   728.8	& $	 -3.21 \times 10^{-6}	$ & $	 -1.50 \times 10^{5}	$ & $	-	 $ \\
$	$	&	7/2	&	1	&	-1	&	   740.9	& $	 -1.54 \times 10^{-5}	$ & $	 -1.31 \times 10^{5}	$ & $	  4.40 \times 10^{-1}	 $ \\
$	$	&	5/2	&	1	&	0	&	   764.3	& $	 -5.69 \times 10^{-1}	$ & $	 -1.21 \times 10^{5}	$ & $	-	 $ \\
$	$	&	5/2	&	0	&	0	&	   816.5	& $	 -2.37 \times 10^{0}	$ & $	 -4.30 \times 10^{0}	$ & $	-	 $ \\
$	$	&	5/2	&	0	&	0	&	   888.9	& $	 -6.37 \times 10^{1}	$ & $	 -2.70 \times 10^{1}	$ & $	-	 $ \\
$	$	&	5/2	&	0	&	0	&	   943.3	& $	 -2.03 \times 10^{0}	$ & $	 -6.10 \times 10^{1}	$ & $	-	 $ \\
$\ket{1/2,	1/2,	3,	2}$	&	5/2	&	1	&	1	&	   704.5	& $	 -1.79 \times 10^{1}	$ & $	 -9.94 \times 10^{4}	$ & $	  1.70 \times 10^{-1}	 $ \\
$	$	&	7/2	&	1	&	0	&	   734.6	& $	-	$ & $	-	$ & $	-	 $ \\
$	$	&	5/2	&	1	&	1	&	   760.4	& $	 -6.03 \times 10^{-1}	$ & $	 -1.33 \times 10^{5}	$ & $	  1.20 \times 10^{1}	 $ \\
$	$	&	9/2	&	1	&	-1	&	   773.1	& $	 -1.65 \times 10^{-6}	$ & $	 -1.32 \times 10^{5}	$ & $	-	 $ \\
$	$	&	7/2	&	1	&	0	&	   798.3	& $	 -1.89 \times 10^{-6}	$ & $	 -1.22 \times 10^{5}	$ & $	  8.67 \times 10^{-1}	 $ \\
$	$	&	5/2	&	1	&	1	&	   824.7	& $	 -1.39 \times 10^{-3}	$ & $	 -1.17 \times 10^{5}	$ & $	  7.87 \times 10^{-1}	 $ \\
$	$	&	5/2	&	0	&	0	&	   896.2	& $	 -1.39 \times 10^{2}	$ & $	 -2.07 \times 10^{1}	$ & $	  7.23 \times 10^{-1}	 $ \\
$	$	&	5/2	&	0	&	0	&	   939.6	& $	 -2.00 \times 10^{0}	$ & $	 -9.08 \times 10^{1}	$ & $	  2.10 \times 10^{1}	 $ \\
$	$	&	5/2	&	0	&	0	&	  1019.1	& $	 -1.30 \times 10^{-3}	$ & $	 -5.03 \times 10^{1}	$ & $	  7.55 \times 10^{-1}	 $ \\
$\ket{1/2,	-1/2,	3,	2}$	&	3/2	&	1	&	1	&	   694.8	& $	 -3.90 \times 10^{-1}	$ & $	 -6.86 \times 10^{4}	$ & $	-	 $ \\
$	$	&	5/2	&	1	&	0	&	   728.5	& $	-	$ & $	-	$ & $	-	 $ \\
$	$	&	3/2	&	1	&	1	&	   750.1	& $	 -1.75 \times 10^{1}	$ & $	 -1.00 \times 10^{5}	$ & $	  1.48 \times 10^{-1}	 $ \\
$	$	&	7/2	&	1	&	-1	&	   761.5	& $	-	$ & $	-	$ & $	-	 $ \\
$	$	&	5/2	&	1	&	0	&	   784.8	& $	-	$ & $	-	$ & $	-	 $ \\
$	$	&	3/2	&	1	&	1	&	   811.2	& $	 -5.68 \times 10^{-1}	$ & $	 -1.30 \times 10^{5}	$ & $	  1.28 \times 10^{1}	 $ \\
$	$	&	7/2	&	1	&	-1	&	   828.0	& $	 -2.27 \times 10^{-6}	$ & $	 -1.28 \times 10^{5}	$ & $	  8.42 \times 10^{-1}	 $ \\
$	$	&	5/2	&	1	&	0	&	   853.8	& $	 -1.33 \times 10^{-6}	$ & $	 -1.20 \times 10^{5}	$ & $	  8.37 \times 10^{-1}	 $ \\
$	$	&	3/2	&	0	&	0	&	   854.3	& $	  1.43 \times 10^{0}	$ & $	  9.74 \times 10^{0}	$ & $	-	 $ \\
$	$	&	3/2	&	0	&	0	&	   941.6	& $	 -1.33 \times 10^{2}	$ & $	 -2.17 \times 10^{1}	$ & $	  6.01 \times 10^{-1}	 $ \\
$	$	&	3/2	&	0	&	0	&	   989.9	& $	 -1.89 \times 10^{0}	$ & $	 -8.42 \times 10^{1}	$ & $	  2.12 \times 10^{1}	 $ \\

\end{tabular}
\end{ruledtabular}
\end{table*}
\end{widetext}

\bibliography{Bib_LiCsFeshbachResonances}

\end{document}